\documentclass[twocolumn,secnumarabic,,amsmath,amssymb, nobibnotes, aps, prd, superscriptaddress]{revtex4-2}

\usepackage{amsmath}
\usepackage{amssymb}
\usepackage{graphicx}
\usepackage{dcolumn}
\usepackage{floatrow}
\usepackage{xcolor}

\begin{document}
\title{Twisted two-dimensional material stacks for polarization optics}
\author{Kaveh Khaliji}
\email{khali161@umn.edu}
\affiliation{Department of Electrical and Computer Engineering, University of Minnesota, Minneapolis, Minnesota 55455, USA}
\author{Luis Mart\'{\i}n-Moreno}
\affiliation{Instituto de Nanociencia y Materiales de Aragon (INMA), CSIC-Universidad de Zaragoza, Zaragoza 50009, Spain}
\affiliation{Departamento de Fisica de la Materia Condensada, Universidad de Zaragoza, Zaragoza 50009, Spain}
\author{Phaedon Avouris}
\affiliation{Department of Electrical and Computer Engineering, University of Minnesota, Minneapolis, Minnesota 55455, USA}
\affiliation{IBM T. J. Watson Research Center, Yorktown Heights, New York 10598, USA}
\author{Sang-Hyun Oh}
\affiliation{Department of Electrical and Computer Engineering, University of Minnesota, Minneapolis, Minnesota 55455, USA}
\author{Tony Low}
\email{tlow@umn.edu}
\affiliation{Department of Electrical and Computer Engineering, University of Minnesota, Minneapolis, Minnesota 55455, USA}

\begin{abstract}
The ability to control light polarization state is critically important for diverse applications in information processing, telecommunications, and spectroscopy. Here, we propose that a stack of anisotropic van der Waals materials can facilitate the building of optical elements with Jones matrices of unitary, Hermitian, non-normal, singular, degenerate, and defective classes. We show that the twisted stack with electrostatic control can function as arbitrary-birefringent wave-plate or arbitrary polarizer with tunable degree of non-normality, which in turn give access to plethora of polarization transformers including rotators, pseudorotators, symmetric and ambidextrous polarizers. Moreover, we discuss an electrostatic-reconfigurable stack which can be tuned to operate as four different polarizers and be used for Stokes polarimetry. 
\end{abstract}


\date{\today}
\maketitle

\noindent \textit{Introduction} -- Polarization optics or the science of controlling the polarization state of electromagnetic waves has broad applicability in areas such as polarimetric imaging, biosensing, and optical communication \cite{wolff1997, pierangelo2011, brodsky2006}. The central building blocks in polarization optics are polarization transformers; optical devices which scatter polarized light into a pre-defined polarization state. Broadly speaking, there are two strategies used to build such devices. One way is to cascade transmissive optical elements, where stacking order and relative orientation between the principal axes of the elements determine the polarization output \cite{yariv2007, saleh2019}. For instance a common way to build a circular polarizer is by cascading a linear polarizer and a quarter-wave plate, with the transmission direction of the polarizer at $45^{\circ}$ to the fast axes of the retarder. However, state-of-art quarter-wave plate uses bulky linear birefringent crystals, since it requires a significant propagation distance to establish the phase difference between orthogonal polarizations. The other way is to utilize metasurfaces with patterned metallic or dielectric structures, whose meta-elements can be designed to deliver a pre-selected polarization transformation \cite{yu2014, kim2017, kuwata2005, grady2013, zhao2013, ding2015, li2015, pfeiffer2013}. For example, patterned structures with no rotational symmetry, but with out-of-plane (including surface normal) mirror symmetry would guarantee linear birefringence upon normal illumination \citep{menzel2010, armitage2014}. The built-in spatial symmetries of meta-elements by design restricts the transformation functions these devices can deliver.\\

The family of atomically thin two-dimensional materials which emerged in 2004 with the isolation of graphene \citep{novoselov2004, novoselov2005, zhang2005}, now includes materials with diverse optical properties ranging from dielectrics to metals with anisotropy or hyperbolicity in the terahertz to mid-infrared spectral range \cite{xia2014, wang2020, low2017, basov2016}. In addition to the steady growth in the 2D materials library, there has been great progress on the fabrication of high quality van der Waals heterostructures \cite{geim2013, liu2020d, pizzocchero2016}, control of twist angle between stacked layers \cite{frisenda2018, du2021}, and modulation of Drude weight via doping up to maximum carrier concentration of $10^{14}$ cm$^{-2}$ \cite{rizzo2020}.\\

The previous works on hetero-stack twisted 2D materials has been exclusively on \textit{near-field} polaritonic optics \cite{hu2020, chen2020}. Here, however, we will explore their potential for polarization control in the \textit{far-field}. We focus on non-depolarizing polarization transformers, i.e. those which can be described via Jones matrix \citep{lu1994}. We show that stacking of twisted anisotropic 2D materials, even in its homogenous form without any patterning, allows for facile realization of diverse optical polarization transformers. Herein, both \textit{anisotropy} and \textit{twist} are key ingredients breaking the rotational and mirror symmetries, which in turn allows for a free-form Jones matrix which can be selectively tuned via controlled twisting and stacking order. We theoretically demonstrate how electrostatic doping can give the stack an unprecedented ability to toggle between functionalities via tuning the eigen-spectrum and eigen-polarization of the Jones matrix.\\


\begin{figure*}
\centering
\includegraphics[width=\linewidth]{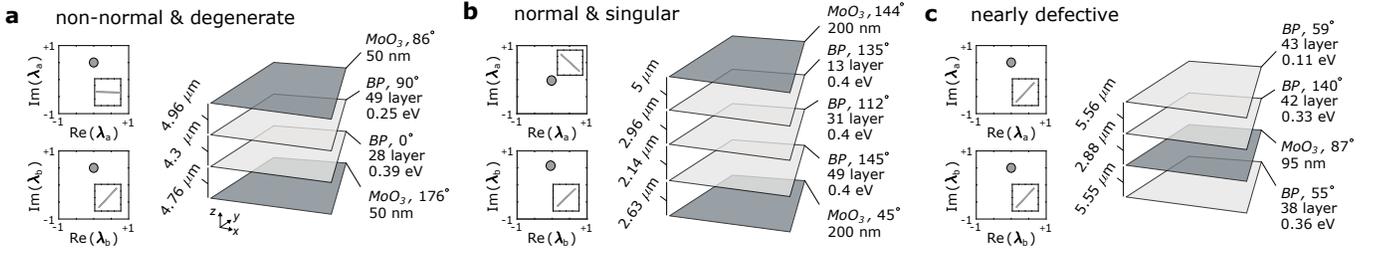}
\caption{The eigen-view of various classes of Jones matrices constructed by twisted stack of BP and MoO$_{3}$ which are represented by light grey and dark grey planes, respectively. In (a)-(c): the left panels depict the two eigenvalues and the corresponding polarization ellipses of the Jones matrices of the structures shown on the right.}
\label{}
\end{figure*}

\noindent  \textit{Mathematical Preliminaries} -- The Jones matrix of an optical element provides a full description of the polarization changes that a beam of light undergoes as it passes through the element. The Jones matrix in its general form can be written as \citep{gutierrezvega2020a, gutierrezvega2020b}:
\begin{equation}
\begin{split}
\textbf{J} = 
\begin{bmatrix}
J_{xx} & J_{xy} \\
J_{yx} & J_{yy}
\end{bmatrix}
= \textbf{P}\,\bf{\Lambda}\,\textbf{P}^{-1}
\end{split}
\label{A1}
\end{equation}
The second equality in \eqref{A1} is the Jordan decomposition, which states that for any Jones matrix there exists an invertible matrix $\textbf{P}$ and a Jordan matrix $\bf{\Lambda}$ which can adopt one of the following forms:
\begin{equation}
\begin{split}
\bf{\Lambda}_{1} = 
\begin{bmatrix}
\lambda_{a} & 0 \\
0 & \lambda_{b}
\end{bmatrix},  
\bf{\Lambda}_{2} = 
\begin{bmatrix}
\lambda & 1 \\
0 & \lambda
\end{bmatrix}
\end{split}
\label{A2}
\end{equation}

The Jordan form $\bf{\Lambda}_{1}$ corresponds to matrices $\textbf{J}$ that are diagonalizable and have two linearly independent eigenvectors $|a\rangle$, $|b\rangle$ and two eigenvalues $\lambda_{a}$, $\lambda_{b}$, respectively. In this case, $\textbf{P}$ is a matrix whose columns are the eigenvectors $|a\rangle$ and $|b\rangle$, with $|a\rangle \neq |b\rangle$. The diagonalizable $\textbf{J}$ is normal or non-normal depending on whether $\langle a | b \rangle = 0$ or not. Note that Hermitian and unitary Jones matrices are normal. The diagonalizable $\textbf{J}$ is degenerate if $\lambda_{a} = \lambda_{b}$ or singular if $\lambda_{a}$ or $\lambda_{b}$ is zero.\\

The Jordan form $\bf{\Lambda}_{2}$ corresponds to defective matrices $\textbf{J}$ that are not diagonalizable. They have only one eigenvector $|a\rangle$ with eigenvalue $\lambda$. In this case, the eigenvector $|a\rangle$ and the so-called generalized eigenvector which satisfies the equation $(\textbf{J} - \lambda \textbf{I})| g \rangle = |a\rangle$ constitute the columns in $\textbf{P}$. Here, $\textbf{I}$ is the identity matrix. Note the defective $\textbf{J}$ can also be singular if $\lambda= 0$. \textit{We use Eqs. \eqref{A1} and \eqref{A2} to construct a desired Jones matrix given its eigenvalues and eigenvectors}. Mathematically, Jones matrix can also be written in the following way (see supplemental info, S1):
\begin{equation}
\begin{split}
\textbf{J} = \frac{\lambda_{a}}{\langle b_{\bot}|a\rangle}|a\rangle \langle b_{\bot}| + \frac{\lambda_{b}}{\langle a_{\bot}|b\rangle}|b\rangle \langle a_{\bot}|
\end{split}
\label{A3}
\end{equation}
for diagonalizable $\textbf{J}$, and
\begin{equation}
\begin{split}
\textbf{J} = \lambda \textbf{I} + \frac{1}{\langle a_{\bot}|g\rangle}|a\rangle \langle a_{\bot}|
\end{split}
\label{A4}
\end{equation}
for defective Jones matrices. Here, $|a_{\bot}\rangle = [-a^{*}_{y}, a^{*}_{x}]^{\text{T}}$ is the orthogonal state to the eigenstate $|a\rangle = [a_{x}, a_{y}]^{\text{T}}$, i.e. $\langle a_{\bot}|a\rangle = 0$. Note that the normal Jones matrix is a special case in Eq. \eqref{A3} with $|b\rangle = |a_{\bot}\rangle$.\\


\noindent  \textit{Designer Jones Matrices via Stack \& Twist} -- Here we consider normal illumination along $z$-axis. The chosen operating frequency of the devices is 20\,THz. The 2D materials we use are black phosphorus (BP) and orthorhombic molybdenum trioxide ($\alpha$-MoO$_{3}$), see supplemental info, S2 for the material parameters \cite{low2014a, low2014b, desousa2017, zheng2019, ma2018, alvarezperez2019}. We assume that the 2D layers are embedded within a uniform dielectric background (air). This rules out asymmetric effects due to the presence of a substrate and dielectric spacers. The scattering coefficients of the $N$-layer stack, where each layer can be either $\alpha$-MoO$_{3}$ or BP are obtained via the transfer-matrix-method (see supplemental info, S3) \cite{kotov2019}. The twist angle for each layer is measured relative to $x$-axis. The design parameters (including thickness and twist angle of BP and $\alpha$-MoO$_{3}$, the layer spacing, and the chemical potential in BP) are determined numerically by use of the nonlinear programming solver FMINCON in MATLAB and minimization of the cost function defined as: $\sum_{ij}|J_{ij} - J_{ij}^{d}|$, where $\{i,j\} \in \{x,y\}$ and $J^{d}$ is the element of desired Jones matrix. We should comment on the choice of BP and MoO$_{3}$. This is because for these materials the analytic conductivity expressions for different thicknesses and dopings are available and one can efficiently solve for the stack which minimizes the cost function.\\


Figure 1 shows examples of Jones matrices in three classes which can be generated from a few layer stack of BP and $\alpha$-MoO$_{3}$. In Fig. 1(a) the structure gives a degenerate $\left( \lambda_{a}=\lambda_{b}=0.5i \right)$ and non-normal $\left( \langle0^{\circ}|45^{\circ}\rangle\neq0 \right)$ Jones matrix. In panel (b), the Jones matrix is normal $\left( \langle-45^{\circ}|45^{\circ}\rangle = 0 \right)$ and singular $\left( \lambda_{a}=0, \lambda_{b}=0.5i\right)$. In panel (c), the stack gives a nearly defective Jones matrix, i.e. the parameter setting points to an exceptional point in polarization space where the eigenvalues and eigenvectors of the Jones matrix nearly coalesce on $0.5i$ and $|45^{\circ}\rangle$, respectively.\\

The stacks in Fig. 1 are examples of static Jones matrix engineering, static in the sense that one Jones matrix corresponds to  one stack. Next, we focus on electrostatic doping and its ability to tune the eigen-vectors and eigenvalues of a few polarization transformers. More explicitly, we discuss transformers in which the layer number, layer separations, and twist angles are fixed and we only change the doping in the BP layers \cite{peng2017, chaves2021}. Note that the MoO$_{3}$ optical response is due to its optical phonons and is not tunable with doping. We will further show that with a increase in complexity, it is possible to have a single system which presents several (doping dependent) Jones Matrix functionalities. Additionally, one can use electrostatic control to retrieve the desired functionality in the case when twist angles or layer thicknesses deviate from the optimum settings (see supplemental info, S4).\\

To better guide the reader, here we outline what lays ahead: (1) We begin with tunable wave-plates. We describe how the Jones matrix of a generic wave-plate looks like (unitary). We show that the electrostatic control allows for tuning both the wave-plate retardation (its eigenvalues) and birefringence type (its eigenvectors). (2) We discuss polarizers next and introduce the Jones matrix of a generic polarizer (Hermitian and singular). We will use these polarizers to build a Stokes polarimeter. The twisted stack can function as four different polarizers, where one can switch from one polarizer to the other via change in the BP doping. (3) We close with a discussion on non-normal and ambidextrous polarizers. The first corresponds to non-normal and singular while the other corresponds to defective and singular Jones matrices. These polarizers are peculiar in a sense that they are asymmetric, i.e. they act differently depending on which direction light propagates through the stack.\\

\begin{figure}
\centering
\includegraphics[width=0.95\linewidth]{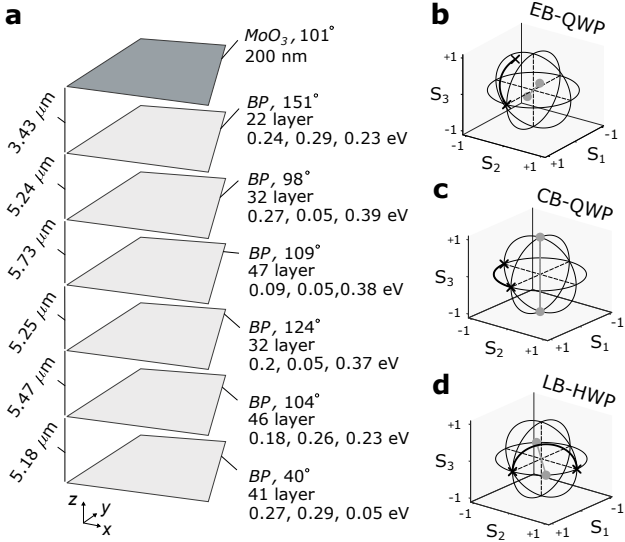}
\caption{(a) The 7-layer BP and MoO$_{3}$ stack to achieve (b) elliptical-birefringent QWP, (c) circular-birefringent QWP, and (d) linear-birefringent HWP. The Fermi energies for each constituent BP multilayer are listed in the order which gives the result in panels (b), (c), and (d). In (b)-(d), the circles denote the eigen-states while crosses represent $|0^{\circ}\rangle$ and ${\bf{J_{R}}}|0^{\circ}\rangle$ to showcase the waveplate-induced rotations in the Poincar\'e sphere.}
\label{}
\end{figure}

\begin{figure}[!b]
\centering
\includegraphics[width=0.95\linewidth]{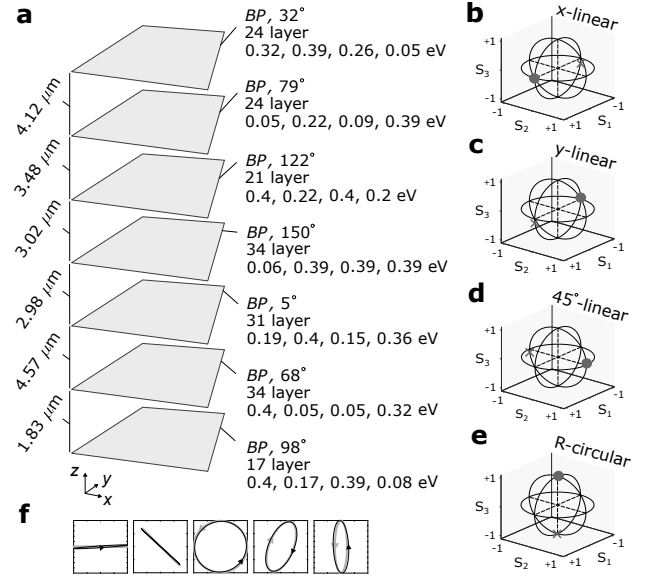}
\caption{(a) The 7-layer twisted BP stack can switch between (b) $x$-linear polarizer, (c) $y$-linear polarizer, (d) $45^{\circ}$-linear polarizer, and (e) right-handed circular polarizer. The Fermi energies for each constituent BP multilayer are listed in the order which gives the result in panels (b), (c), (d), and (e). In (b)-(e), the crosses represent the eigen-polarization with zero eigenvalue. (f) The polarization ellipse corresponding to input light (grey) and those measured (black) using the stack in panel (a) as Stokes polarimeter. From left to right, the input Jones vectors are $[1,0]^{\text{T}}$, $[0.7071,-0.7071]^{\text{T}}$, $[0.7071,0.7071i]^{\text{T}}$, $[0.5, 0.433+0.75i]^{\text{T}}$, and $[0.2236,0.9747i]^{\text{T}}$.}
\label{}
\end{figure}

\begin{figure*}
\floatbox[{\capbeside\thisfloatsetup{capbesideposition={right,center},capbesidewidth=5.5cm}}]{figure}[1.3\FBwidth]{\caption{(a) The 7-layer twisted BP stack optimized to function as two asymmetric polarizers with (b) $\Theta = 90^{\circ}$ and (c) $\Theta\sim 0^{\circ}$. The eigen-polarizations in (b) are $|45^{\circ}\rangle$ and $|\text{R}\rangle$. In (c) the eigen-polarizations are set to coalesce on $|45^{\circ}\rangle$. In (b) and (c), the crosses represent the eigen-polarization with zero eigenvalue. The Fermi energies for each constituent BP multilayer are listed in the order which gives the result in panels (b) and (c). In (d) and (e), the action of the non-normal polarizer in (b) and the nearly ambidextrus polarizer in (c) are visualized in forward direction, respectively. (f) and (g) are the same as (d) and (e) for backward propagation. In (d)-(g) black arrows show the propagation direction and black crosses denote blocked transmission.}\label{}}
{\includegraphics[trim = 0mm 0mm 0mm 0mm, clip, width=1.0\linewidth]{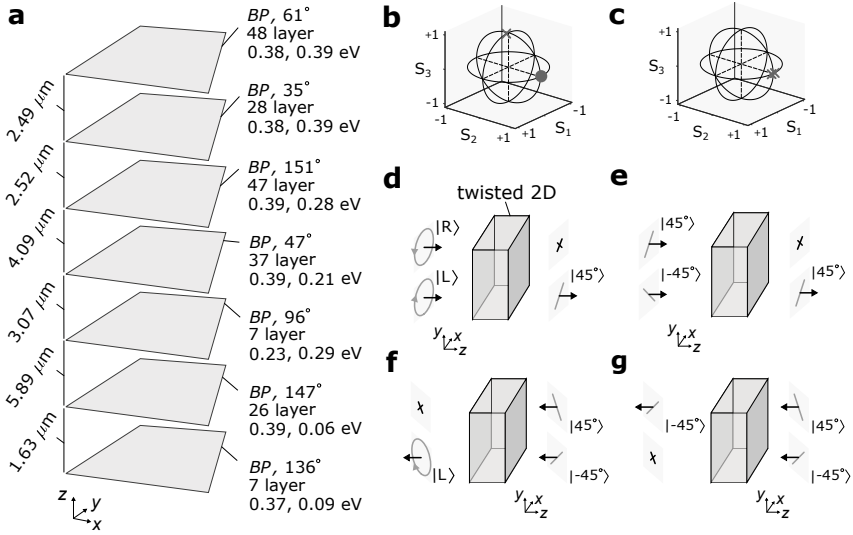}}
\end{figure*}

\noindent \textit{Arbitrary\,-\,Birefringent Wave Plates} -- The term wave plate (retarder) refers to birefringent optical elements where the element anisotropy induces a phase shift $\eta$ between its two orthogonal eigen-polarizations. The most general case is an elliptic retarder, whose eigenstates are given by a pair of orthogonal Jones vectors,
\begin{equation}
\begin{split}
| a \rangle
=
\begin{bmatrix}
\cos\alpha \\
\sin\alpha e^{i\delta}
\end{bmatrix},~~
| b \rangle
=
\begin{bmatrix}
-\sin\alpha e^{-i\delta}\\
\cos\alpha
\end{bmatrix} 
\end{split}
\label{A5}
\end{equation}
corresponding to eigenvalues, $\lambda_{a}=e^{i\eta/2}$ and $\lambda_{b}=e^{-i\eta/2}$, respectively. The $(\alpha, \delta)$ pair determines the orientation $\phi$, $\phi + \pi/2$ and ellipticity angles $\epsilon$, $-\epsilon$ for $|a\rangle$ and $|b\rangle$, respectively, such that:
$\tan 2\phi = \tan 2\alpha \cos\delta$, $\sin2\epsilon = \sin 2\alpha \sin\delta$ \cite{damask2004}. A generic retarder is then represented by a unitary Jones matrix $\bf{J_{R}}$:
\begin{equation}
\begin{split}
{\bf{J_{R}}}(\alpha,\delta,\eta)
= \begin{bmatrix}
c^{2}_{\alpha} e^{i\eta/2} + s^{2}_{\alpha} e^{-i\eta/2} && 
i s_{2\alpha}s_{\eta/2} e^{-i\delta} \\
i s_{2\alpha}s_{\eta/2} e^{i\delta} &&
s^{2}_{\alpha} e^{i\eta/2} + c^{2}_{\alpha} e^{-i\eta/2}
\end{bmatrix} 
\end{split}
\label{A6}
\end{equation}
where, $s_{x}\equiv\sin x$ and $c_{x}\equiv\cos x$. Note that $\delta=0$ denotes linear retarders. $\alpha=45^{\circ}$ and $\delta=\pm90^{\circ}$ gives circular retarders which can also be interpreted as $\eta/2$ polarization rotator i.e. it only rotates the major axis of the polarization ellipse with angle $\eta/2$ but keeps the ellipticity and handedness intact. The retarder action on the Poincar\'e sphere can be described as a rotation with the rotation axis defined by the points corresponding to $|a\rangle$ and $|b\rangle$, i.e. eigen-polarization of ${\bf{J_{R}}}$ and its rotation angle with retardance $\eta$.\\


In Fig. 2(a) we show an example of a tuneable waveplate. In Fig. 2(b), the stack operates as a quarter-wave plate (QWP, $\eta = 90^{\circ}$) with elliptic birefringence characterized by $\alpha = 27.4^{\circ}, \delta = 45^{\circ}$. In Figs. 2(c) and (d) we show, by modifying the electrostatic doping the same structure can also function as a circular or linear birefringent (CB or LB) retarder. The CB-QWP is basically a $45^{\circ}$-polarization rotator. The LB half-wave plate (HWP) with $\alpha=22.5^{\circ}$ corresponds to a $45^{\circ}$-pseudorotator. The latter produces an improper rotation, i.e. it rotates the polarization ellipse major axis by $2\alpha  = 45^{\circ}$ and reverses the ellipse handedness. The LB-HWP can be used to implement right-left circular polarization conversion. Note that the structure in Fig. 2 has transmission efficiency of $\sim$\,0.5 for its two eigen-polarizations. This suggests that the polarization effect is primarily birefringence and not dichroism. The non-unity transmission also indicates that the corresponding Jones matrices are scaled unitary, i.e. ${\bf{J}}{\bf{J}}^{\dag} \propto {\bf{I}}$. We note that the wave-plate functionalities in Fig. 2 can be implemented separately using stacks with fewer layers (see supplemental info, S5).”\\ 



\noindent \textit{Stokes Polarimetry \& Asymmetric Polarizer} -- A diattenuator refers to an optical element that exhibits anisotropic intensity attenuation. The most general case of a diattenuator is the elliptic diattenuator (or elliptic partial polarizer), whose eigenstates are given by a pair of orthogonal Jones vectors $| a \rangle$ and $| b \rangle$ as defined in Eq. \eqref{A5} corresponding to real eigenvalues $\lambda_{a}=p_{1}$ and $\lambda_{b}=p_{2}$, respectively, with $0 \leq p_{1,2} \leq 1$. A generic diattenuator is then represented by a Hermitian Jones matrix $\bf{J_{D}}$:
\begin{equation}
\begin{split}
{\bf{J_{D}}}(\alpha,\delta,p_{1},p_{2})
=
\begin{bmatrix}
p_{1}c^{2}_{\alpha} + p_{2}s^{2}_{\alpha} && 
s_{\alpha}c_{\alpha}(p_{1}-p_{2}) e^{-i\delta} \\
s_{\alpha}c_{\alpha}(p_{1}-p_{2}) e^{i\delta} &&
p_{1}s^{2}_{\alpha} + p_{2}c^{2}_{\alpha}
\end{bmatrix} 
\end{split}
\label{A7}
\end{equation}
where, $s_{x}\equiv\sin x$ and $c_{x}\equiv\cos x$. When $p_{2}=0$, the diattenuator totally extinguishes the eigenstate $|b\rangle$ and is called an elliptic polarizer.\\

Here we show that the cascaded twisted 2D materials can be used for Stokes polarimetry, i.e. to determine the unknown polarization state of the input light. The polarimetry is achieved by electrically toggling between 4 distinct polarizers with different doping configurations as light passes through a 7-layer twisted BP stack (see Fig. 3). The doping can tune the polarizer eigen-vectors, as it allows the stack to switch between linear and circular polarizers. We can also tune the polarizer eigen-values, while not changing its eigen-vectors. This is shown in Figs. 3(b) and (c), where the stack can be switched between $x$- and $y$-linear polarizers. These so-called cross polarizers have similar eigen-vectors, but have the zero eigenvalue swapped. The eigen-vectors corresponding to non-zero eigenvalues are used to accurately recover the input polarization state via  polarimetric data reduction equation (see supplemental info, S6) \cite{tyo2006}.\\

Till now, we have discussed only polarizers resulting from normal  Jones matrices. These are symmetric polarizers i.e. they project any input into one output polarization, in both forward and backward propagation directions \cite{pfeiffer2014}. To infer this, one should note that backward and forward Jones matrices are related: $\textbf{J}^{b} = \textbf{J}^{\text{T}}$ \cite{shi2020}. This together with Eq. \eqref{A3} gives ${\bf{J}} \propto |a\rangle \langle a|$ and ${\textbf{J}}^{b} \propto |a^{b}\rangle \langle a^{b}|$, where $|a\rangle = [a_{x}, a_{y}]^{\text{T}}$ and $|a^{b}\rangle  = [a^{*}_{x}, a^{*}_{y}]^{\text{T}}$ denote identical polarization states in forward and backward propagation directions (see supplemental info, S7) \cite{fedotov2006, menzel2010b}.\\  

Broadly speaking, the polarizer action can also be achieved via non-normal or defective singular Jones matrices \cite{tudor2016}. These matrices represent asymmetric polarizers. Let us illustrate this for the case of a defective Jones matrix, where Eq. \eqref{A4} gives ${\bf{J}} \propto |a\rangle\langle a_{\bot}|$. Note the eigenvector of this so-called ambidextrous polarizer corresponds to an eigenvalue of zero, i.e. $|a\rangle$ the eigenstate of the optical element is blocked. The corresponding backward Jones matrix can be written as ${\bf{J}}^{b} \propto |a^{b}_{\bot}\rangle\langle a^{b}|$. Note the blocked eigenstate of ${\bf{J}}^{b}$ is $|a^{b}_{\bot}\rangle$. \\

In Fig. 4 we show that a 7-layer BP stack, via electrostatic doping, can function as both non-normal asymmetric polarizer $|45^{\circ}\rangle\langle \text{L}|$ and nearly ambidextrous polarizer $|45^{\circ}\rangle\langle-45^{\circ}|$. From Figs. 4(b) and (c), it is clear that the doping can tune the angle between polarizer eigen-vectors, i.e. its degree of non-normality. The latter is measured with $\Theta$, the angle subtended between the polarizer eigenstates on the Poincar\'e sphere. $\Theta=0^{\circ}$ and $180^{\circ}$ denote ambidextrous and normal polarizers respectively. In Figs. 4(d)-(g) we highlight the asymmetric operation of the non-normal and ambidextrous polarizers. Note that the non-normal polarizer acts as a linear polarizer in forward and a circular polarizer in backward directions, while the nearly ambidextrous polarizer has orthogonal linear polarization states as its outputs in both forward and backward directions.\\


\noindent \textit{Concluding Remarks} -- Our results concretely demonstrate that it is possible to control the Jones matrix entries of such stratified structures by adjusting the doping, twist angle, and stacking order of anisotropic 2D layers. Note that the two material systems included in this work do not allow the access to all possible signs of conductivity components changes in the mid-IR. This restrict the input-output polarization mapping that can be assessed by our BP-MoO$_{3}$ structure. However, using new materials including metallic 2D stacks with larger conductivity and even patterned structures, which allow access to their plasmons, clearly suggest twisted hetero-2D stacks constitute a solid choice to build electrostatic-reconfigurable polarization transformers.\\

\noindent \textit{Acknowledgments} -- K. K. and T. L. acknowledge partial support by the National Science Foundation, NSF/EFRI Grant No. EFRI-1741660, and the DDF UMN support for K. K. LM-M acknowledges Project PID2020-115221GB-C41 financed by MCIN/AEI/10.13039/501100011033 and the Aragon Government through Project Q-MAD. S.-H. O. acknowledges support from the Samsung Global
Research (GRO) Program and the Sanford P. Bordeau Chair at the University of Minnesota.


\bibliography{./refs_poloptics}

\end{document}